\documentclass{jpconf}
\usepackage{citesort}
\usepackage{amsmath}
\usepackage{amsfonts}
\usepackage{amssymb}
\usepackage{graphicx}

\newcommand{\be}{\begin{eqnarray}}
\newcommand{\ee}{\end{eqnarray}}
\newcommand{\nn}{\nonumber}

\def \TeV{\mathop{\mbox{\normalfont TeV}}\nolimits}
\def \Br{\mathop{\mbox{\normalfont Br}}\nolimits}
\def \dL{\mathcal{L}}

\def \GeV{\mathop{\mbox{\normalfont ~GeV}}\nolimits}

\def\half{{\textstyle{1 \over 2}}}

\providecommand{\keywords}[1]{\textit{#1}}

\begin{document}

\title{Higgs decay mediated by top-quark with flavor-changing neutral scalar interactions}

\author{J. A. Orduz-Ducuara}

\address{
Divisi\'on de Matem\'aticas e Ingenier\'ia,\\
FES-Acatl\'an, UNAM\\
C.P. 53150, \\
Estado de M\'exico, M\'exico.
}

\ead{jaorduz@ciencias.unam.mx
}

\begin{abstract}
We explore the flavor-changing 
parameters mediated by a Higgs boson within 
the THDM-III context. 
In particular, 
the $h\to t^* c$ 
processes, and check the high suppression 
for the FC 
in the THDM-III context for the low $t_\beta$ parameters.
Our exploration in the $\chi_{ij}^{u}-\chi_{ij}^{d}-$parameter space shows the 
allowed regions for different 
$t_\beta^{}$ values. 
We explored different modes for 
Higgs decays, considered the experimental constraints 
to get scattering plots for the FC parameters and 
some relevant decay modes.
We expect future results to figure out the 
FC and its implications in the scalar sector.
\end{abstract}

\section{Introduction}

Different experiments 
have examined the Standard Model 
observables which have good agreement 
with the theoretical results.
Currently LHC explores the nature at energy 
scale of order  $\TeV'$s and the SM works 
very well to analyze the structure 
of the matter. 
However we have some questions 
to solve: 
the matter-antimatter asymmetry, 
the CP violation and 
flavor-changing neutral currents (FCNC) mediated by gauge 
and scalar bosons. In this document, 
we shall discuss the last one where 
the neutral scalar boson can change 
the fermion flavor. 

We shall discuss the Flavor-Changing Neutral 
Scalar Interactions (FCNS) in the Two-Higgs 
Doublet Model type III (THDM-III) context  considering different 
scenarios for the parameters of the model. 
Previous studies have considered 
different versions for this THDM-III 
\cite{Arroyo:2013tna}. In fact, the total Higgs decay 
is given by \cite{DiazCruz:2012xc}: 
\be
\Gamma_{total} \approx
\Gamma{(h\to b\bar{b})}
+ \Gamma{(h\to c\bar{c})}
+ \Gamma{(h\to WW^*)}
+ \Gamma{(h\to ZZ^*)}
+ \Gamma{(h\to c\bar{b} W^-)};\label{eq:gammatotal}
\ee 
however the purpose of this paper is focused on flavor-changing 
mediated by Higgs bosons, in particular, we explore the 
$hct-$vertex, which is interesting for the last term in eq. 
 \eqref{eq:gammatotal}. 
 
Several reports have shown the following 
theoretical and experimental limit values for the 
branching ratios (${\Br}$) \cite{TheATLAScollaboration:2013nia}:
$ \Br{(t \to uh)} = 5.5\times 10^{-6},$ 
$ \Br{(t\to ch)} =1.5\times 10^{-3},$  
$ \Br{(t\to u \gamma)}  \sim10^{-6},$ 
$ \Br{(t\to c\gamma)}  \sim10^{-6},$ 
$ \Br{(t\to uZ)}  \sim10^{-7},$ 
$ \Br{(t\to cZ)}  \sim10^{-7}.$ 

Our calculation introduces the  
Passarino-Veltman functions implemented on 
Looptools, FeynCalc, FeynArts and SARAH 
\cite{Hahn:1998yk,Hahn:2000kx,Staub:2015kfa}.

This letter is organized as follows:
section \ref{sec:models-methods} 
shows the models and methods in the THDM-III context, 
section \ref{sec:results} we expose a results for the 
$h \to t^* c$ process and 
section \ref{sec:conclusions} contains the conclusions.

\section{Models and Methods
\label{sec:models-methods}}

We shall consider the THDM-III  where the most 
general potential is \cite{Gunion:2002zf}
\be
V(\Phi_1 \Phi_2) &=&
m_{11}^2\Phi_1^\dagger\Phi_1+m_{22}^2\Phi_2^\dagger\Phi_2
-[m_{12}^2\Phi_1^\dagger\Phi_2+{\rm h.c.}]
+\half\lambda_1(\Phi_1^\dagger\Phi_1)^2
+\half\lambda_2(\Phi_2^\dagger\Phi_2)^2
\nn\\
&&
+\lambda_3(\Phi_1^\dagger\Phi_1)(\Phi_2^\dagger\Phi_2)
+\lambda_4(\Phi_1^\dagger\Phi_2)(\Phi_2^\dagger\Phi_1)
+\Big\{\half\lambda_5(\Phi_1^\dagger\Phi_2)^2
+\big[\lambda_6(\Phi_1^\dagger\Phi_1)
\nn\\
&&
+\lambda_7(\Phi_2^\dagger\Phi_2)\big]
\Phi_1^\dagger\Phi_2+{\rm h.c.}\Big\}\,. \nn
\ee

In a general way, the Yukawa sector for the THDM-III 
is given by \cite{DiazCruz:2004pj}
\be\label{eq:Lgrnn-gral-THDM-III}
{\dL}^{THDM-III}_{n} &=& 
\frac{g}{2}
\left(\frac{m_{i}}{m_{W}}\right)\bar{d}_i^{}
\left[
\frac{\cos\alpha}{\cos\beta}\delta_{ij}+\frac{\sqrt{2}\sin(\alpha-\beta)}{g\cos\beta}\left(\frac{m_{W}}{m_{i}}\right)\left(\tilde{Y}_{2}^{d}\right)_{ij}
\right]
d_{j}^{}H^{0} \nonumber\\
 & + & \frac{g}{2}
 \left(\frac{m_{j}^{}}{m_{W}}\right)\bar{d}_{i}^{}
 \left[
 -\frac{\sin\alpha}{\cos\beta}\delta_{ij}^{}+
 \frac{\sqrt{2}\cos(\alpha-\beta)}{g\cos\beta}
 \left(\frac{m_{W}}{m_{i}}\right)
 \left(\tilde{Y}_{2}^{d}\right)_{ij}^{}
 \right] d_{j}^{}h^{0} \nn\\
 & + & \frac{ig}{2}
 \left(\frac{m_{i}}{m_{W}}\right)\bar{d}_{i}^{}
 \left[
 -\tan\beta\delta_{ij}^{}+\frac{\sqrt{2}}{g\cos\beta}
 \left(\frac{m_{W}}{m_{i}}\right)
 \left(\tilde{Y}_{2}^{d}\right)_{ij}^{}
 \right]\gamma^{5}d_{j}^{}A^{0} \nonumber \\
 & + & \frac{g}{2}
 \left(\frac{m_{i}}{m_{W}}\right)\bar{u}_{i}^{}
 \left[
 \frac{\sin\alpha}{\sin\beta}\delta_{ij}^{}+\frac{\sqrt{2}
 \sin(\alpha-\beta)}{g\sin\beta}\left(\frac{m_{W}}{m_{i}}\right)\left(\tilde{Y}_{2}^{u}\right)_{ij}^{}
 \right]u_{j}^{}H^{0} \nn\\
 & + & \frac{g}{2}
 \left(\frac{m_{u}}{m_{W}}\right)\bar{u}_{i}^{}
 \left[
 -\frac{\cos\alpha}{\sin\beta}\delta_{ij}^{}+
 \frac{\sqrt{2}\cos(\alpha-\beta)}{g\sin\beta}
 \left(\frac{m_{W}}{m_{i}}\right)
 \left(\tilde{Y}_{2}^{u}\right)_{ij}^{}
 \right]u_{j}^{}h^{0} \nn \\
  &+ & \frac{ig}{2}
  \left(\frac{m_{u}}{m_{W}}\right)\bar{u}_{i}^{}
  \left[
  -\cot\beta\delta_{ij}^{} + 
  \frac{\sqrt{2}}{g\sin\beta}
  \left(\frac{m_{W}}{m_{i}}\right)
  \left(\tilde{Y}_{2}^{u}\right)_{ij}^{}
  \right]\gamma^{5}u_{j}^{}A^{0}.
\ee

Where the \eqref{eq:Lgrnn-gral-THDM-III} is the Lagrangian 
density for the fermion-fermion-$\phi$; and $\phi = A^0, H^0, h^0$ 
which are the pseudoscalar, heavy Higgs and standard model Higgs, 
and the superscript (zero) labels the  
mass eigenstates.

One can rewrite the 
Yukawa couplings as
$
{\widetilde{Y}}_1^{f} = 
\sqrt{2} 
\frac{M_f^{}}{v_{}^{}\cos\beta} - 
{\widetilde{Y}}_2^{f} ~t_\beta^{}; 
$
because we explore a simple model 
where the flavor-changing is due to the 
$\chi_{f}^{i}$-parameters, and it is 
simple if we reduce the number of those parameters, 
which are associated to the Yukawa couplings; namely 
$Y^{f}_{ij} = \sqrt{2}\frac{\sqrt{m_i^{}m_j^{}}}{v} 
\chi^{f}_{ij}$,
where the $m_{i}^{}$ and $m_j^{}$ are the  fermion 
masses and $\chi^{f}_{ij}$ are the dimensionless parameters that will probe the 
flavor-changing mediated by scalar bosons.

We explore the THDM-III because it is possible to have mixing 
between the fermion flavors at tree level.
For the purpose of this paper, we introduce 
$B \to X_s \gamma$ processes considering 
$\Gamma (B \to X_s^{}\gamma) \simeq 
\Gamma (b \to s^{}_{}\gamma),$ since 
the non-perturbative effects are small \cite{ElKhadra:2002wp}. We consider the constraints 
coming from  
$t\to cV, b\to s\gamma, 
h\to l_i^{} l_j^{}, h \to \gamma Z,$
to show the correlation between 
branching ratios $b\to s \gamma$ and 
$h \to \gamma Z$, and found 
excluded regions (e. g. figs. \ref{label1}-\ref{label2}). 

\begin{figure}[h]\centering
\includegraphics[width=14pc, angle=0]{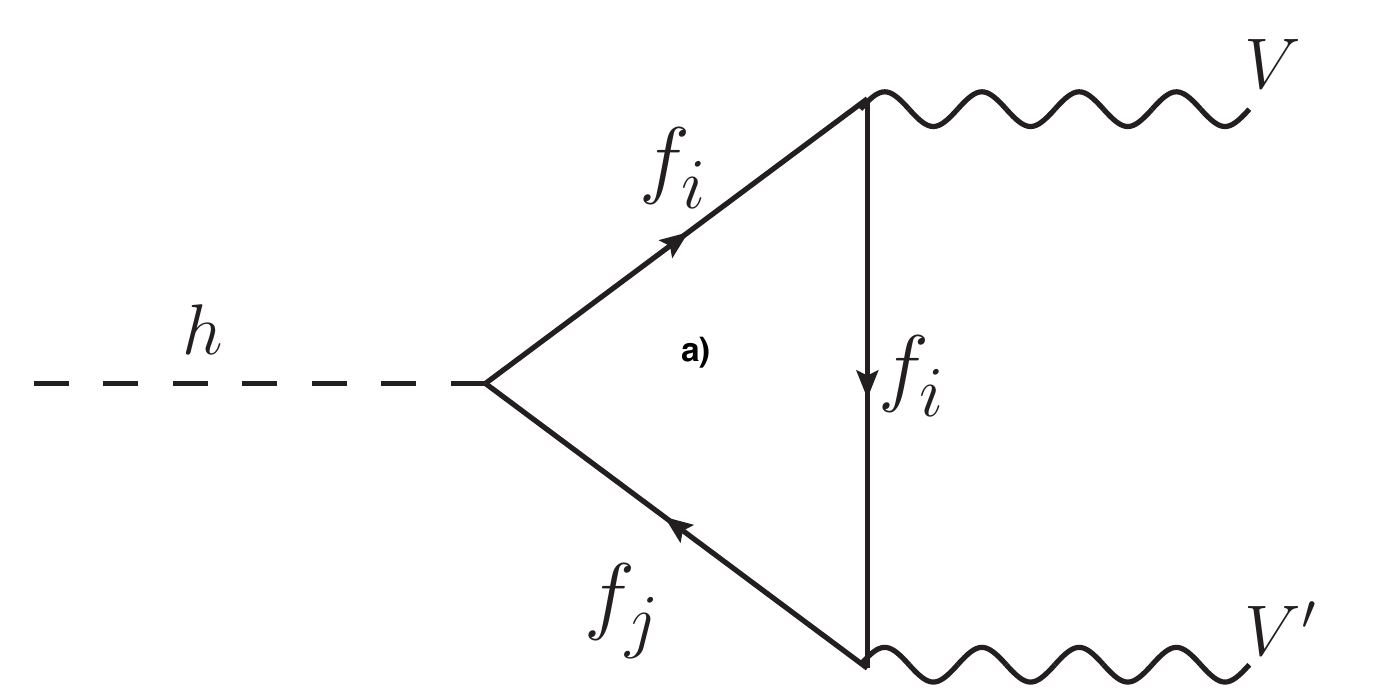}
\includegraphics[width=14pc, angle=0]{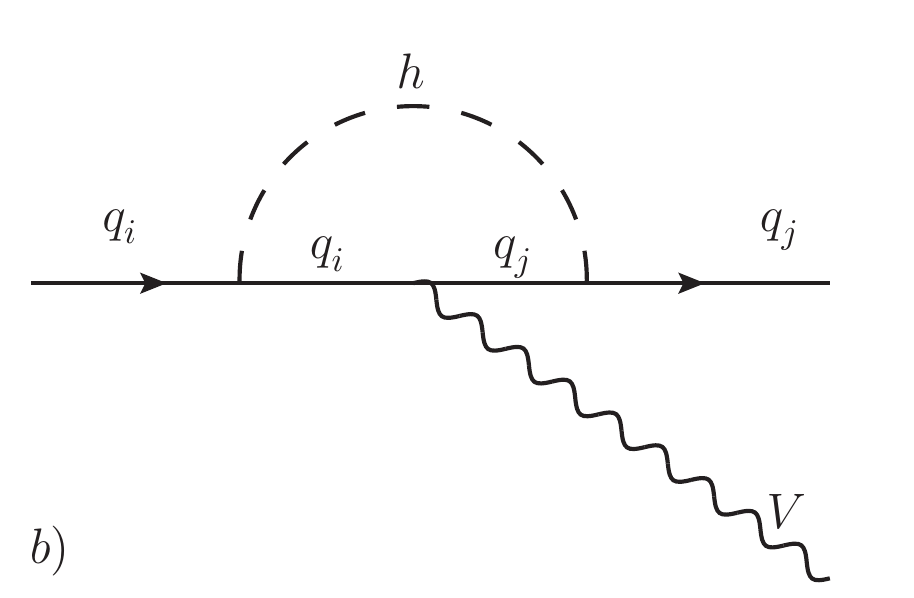}\\
\includegraphics[width=14pc, angle=0]{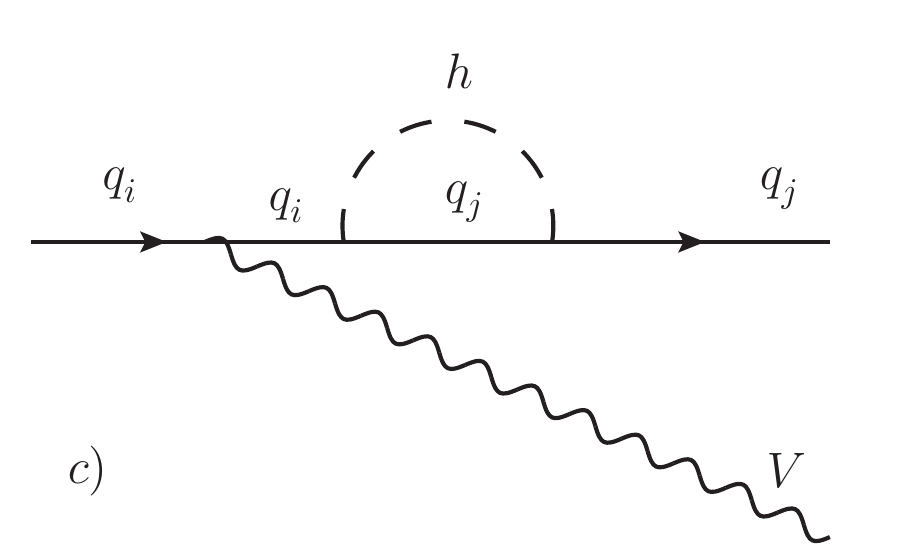}
\includegraphics[width=14pc, angle=0]{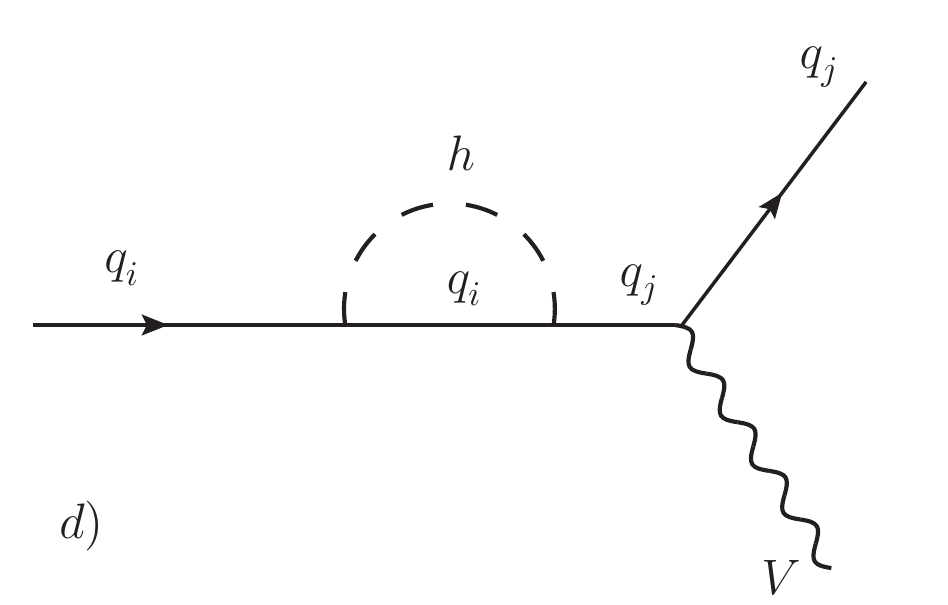}
\caption{
The fig. a) shows the  
Feynman diagrams for $h\to VV'$ considered to constrain 
our parameter analysis.
The figs. c)-d) are the Feynman diagram for the 
 $q_i q_jV$ process at one-loop level with a scalar as a flavor-changing mediator.
}
\end{figure}

\section{Results
\label{sec:results}}

\begin{figure}[h]\centering
\begin{minipage}{14pc}
\includegraphics[width=14pc, angle=-90]{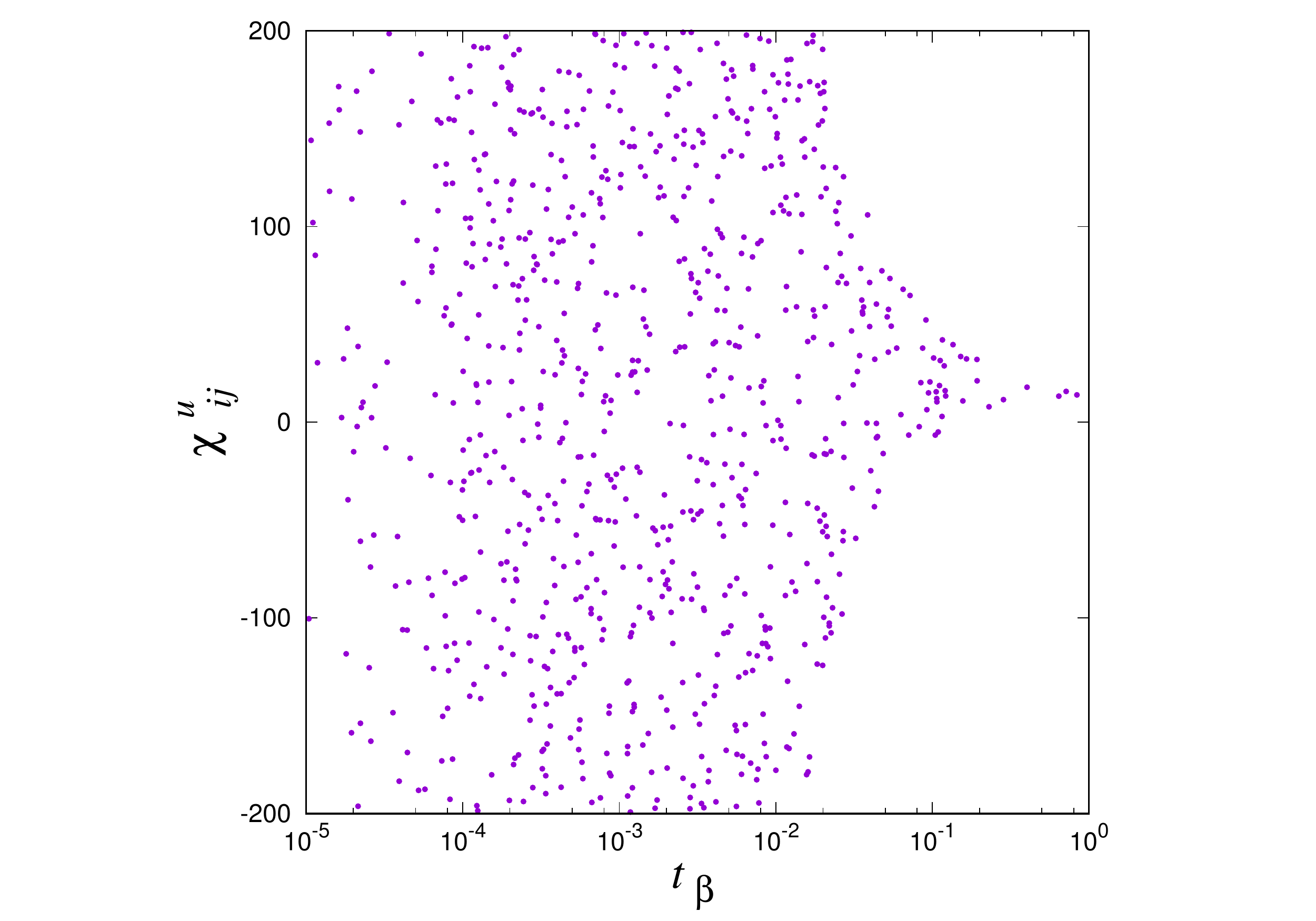}
\caption{\label{label1}Scattering plot for the flavor parameters 
$\chi^{u}_{ij}$versus $t_\beta^{}.$}
\end{minipage}\hspace{2pc}%
\begin{minipage}{14pc}
\includegraphics[width=14pc, angle=-90]{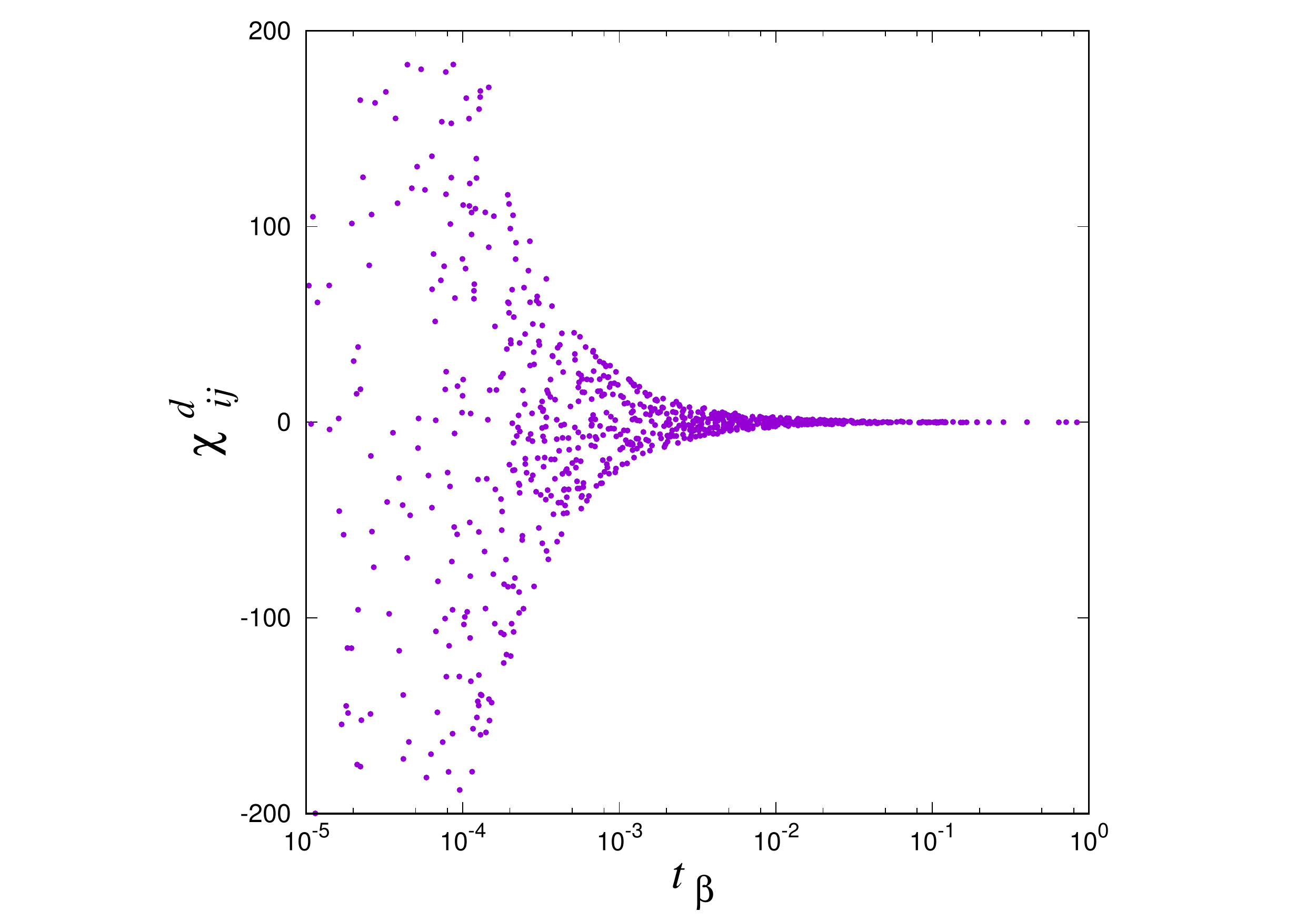}
\caption{\label{label2}Scattering plot for the flavor parameters 
$\chi^{d}_{ij}$ versus $t_\beta^{}.$}
\end{minipage} 
\end{figure}

Figs. \ref{label1} and \ref{label2} 
show a correlation between 
parameters of the model, considering 
different constraints coming from 
$t$ and $b$-quark and $h$ 
decays \cite{Gaitan:2017cfa}. 

\begin{figure}[h]\centering
\includegraphics[scale = 0.5]{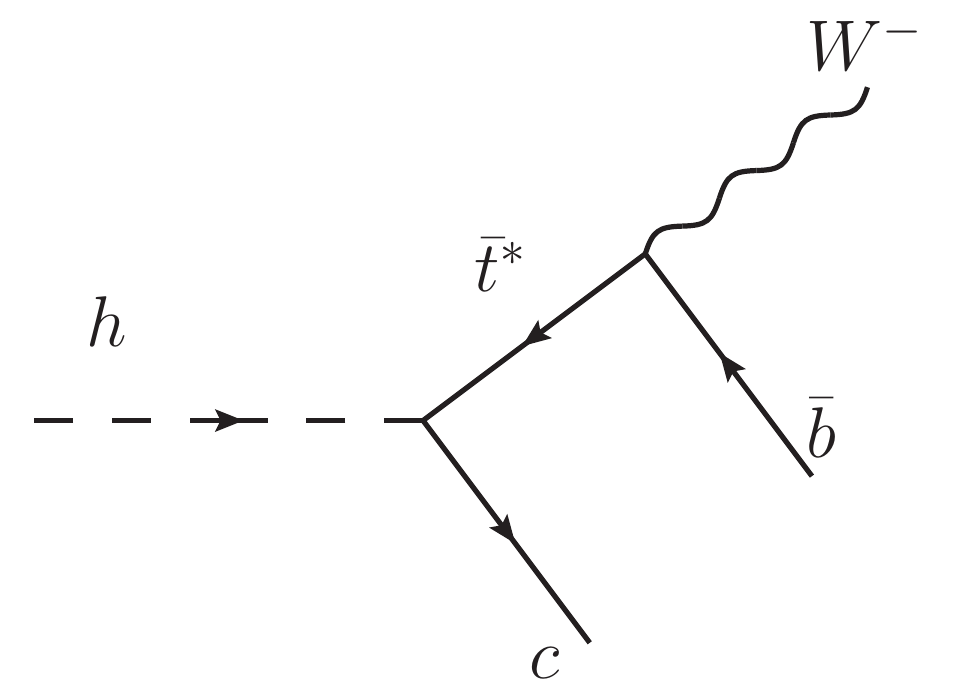}
\caption{\label{fig:FD_h-ts-c}
The Feynman diagram for the 
$h \to t^*_{} c.$}
\end{figure}
The process that is represented in 
fig. \ref{fig:FD_h-ts-c} 
was proposed by ref. \cite{DiazCruz:2012xc} to 
explore the flavor-changing modes.

Our analysis shows that it could be possible to have 
FC if we consider the quark type separately, since 
the parameter space is wider for the u-quark type.
We find that if $t_\beta\lesssim 10^{-3}_{}$ then 
$-200\lesssim\chi_{ij}^{u}\lesssim200,$
while  $t_\beta\lesssim 10^{-4}_{}$ then 
$-200\lesssim\chi_{ij}^{d}\lesssim200.$
Those figs. (\ref{label1} and \ref{label2} ) show that 
$\chi^{d}_{ij}-t_\beta$ is more suppressed than the 
$\chi^{u}_{ij}-t_\beta$  for the lowest $t_\beta$ values.

\begin{figure}[h]\centering
\begin{minipage}{14pc}
\includegraphics[width=14pc, angle=-90]{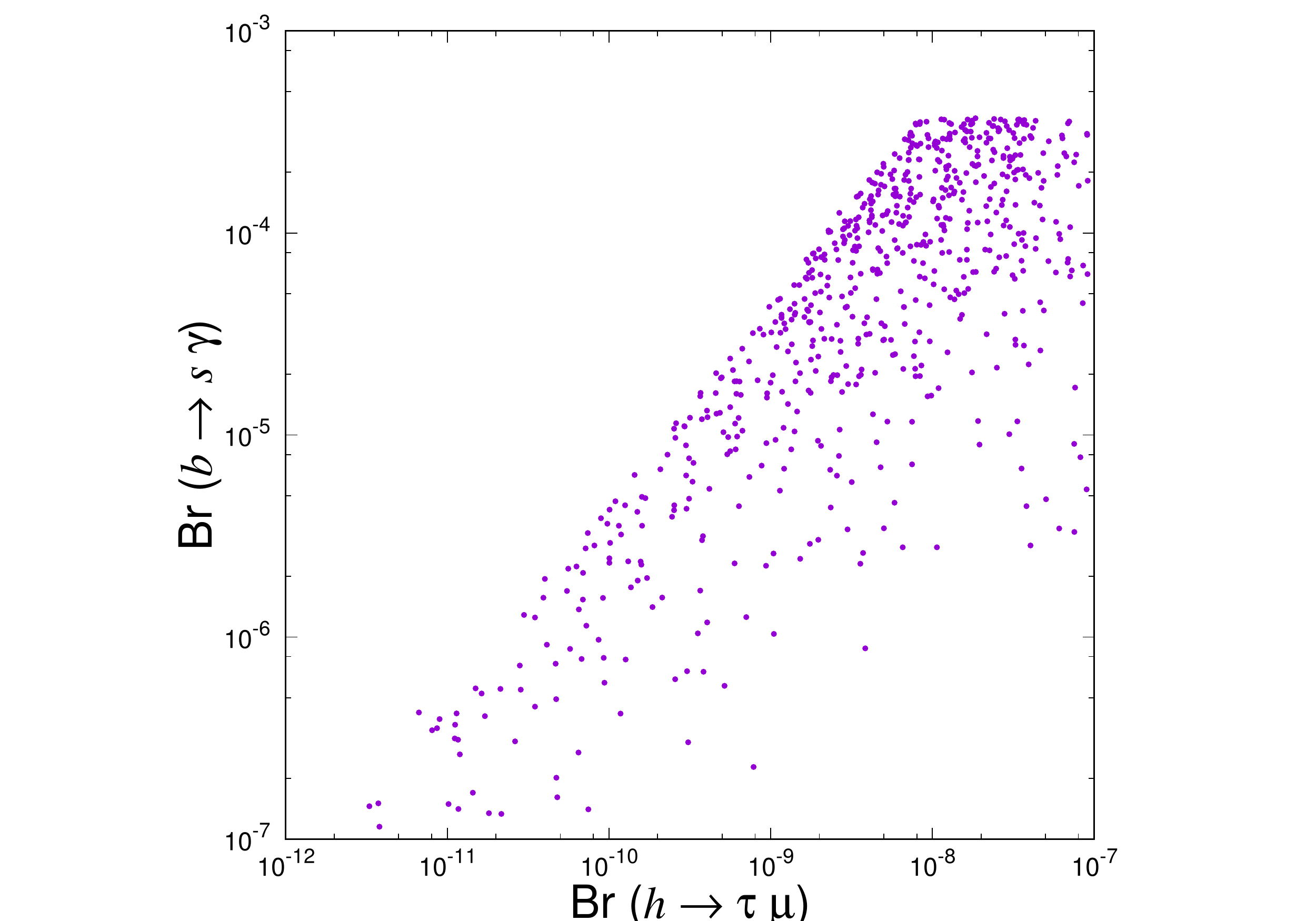}
\caption{\label{label3}
Scattering plot for the branching ratios 
of the processes $b\to s \gamma$ versus 
$h\to \tau\mu.$
}
\end{minipage}\hspace{2pc}%
\begin{minipage}{14pc}
\includegraphics[width=14pc, angle=-90]{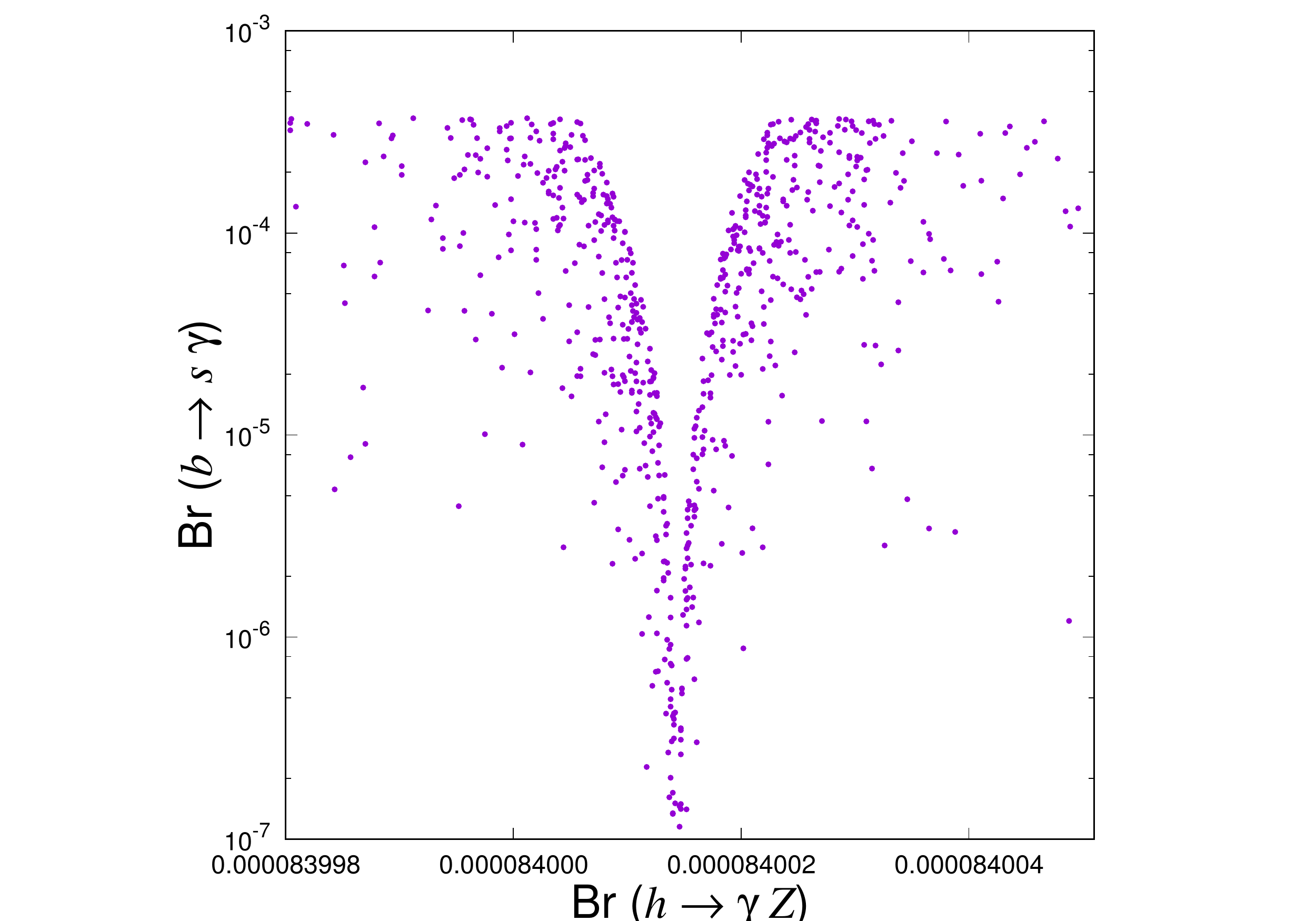}
\caption{\label{label4}
Scattering plot for the branching ratios 
of the processes $b\to s \gamma$ versus 
$h\to \gamma Z.$
}
\end{minipage}
\end{figure}

In figs \ref{label3}-\ref{label4}, 
the dark regions represent the highly allowed region.
We generated randomly the parameter set, as follow,
$-200\leq \chi_{ij}^{u,d}\leq 200, 
0\leq t_\beta^{}\leq 100, 
350~{\GeV}\leq m_{H, H^\pm_{}}^{}\leq 1000~{\GeV},$ 
as well considering the 
experimental bounds for the 
$t, b$ and $h$ branching ratios 
at tree and one-loop level.

\begin{figure}[h]\centering
\includegraphics[scale = 0.6,angle=-90]{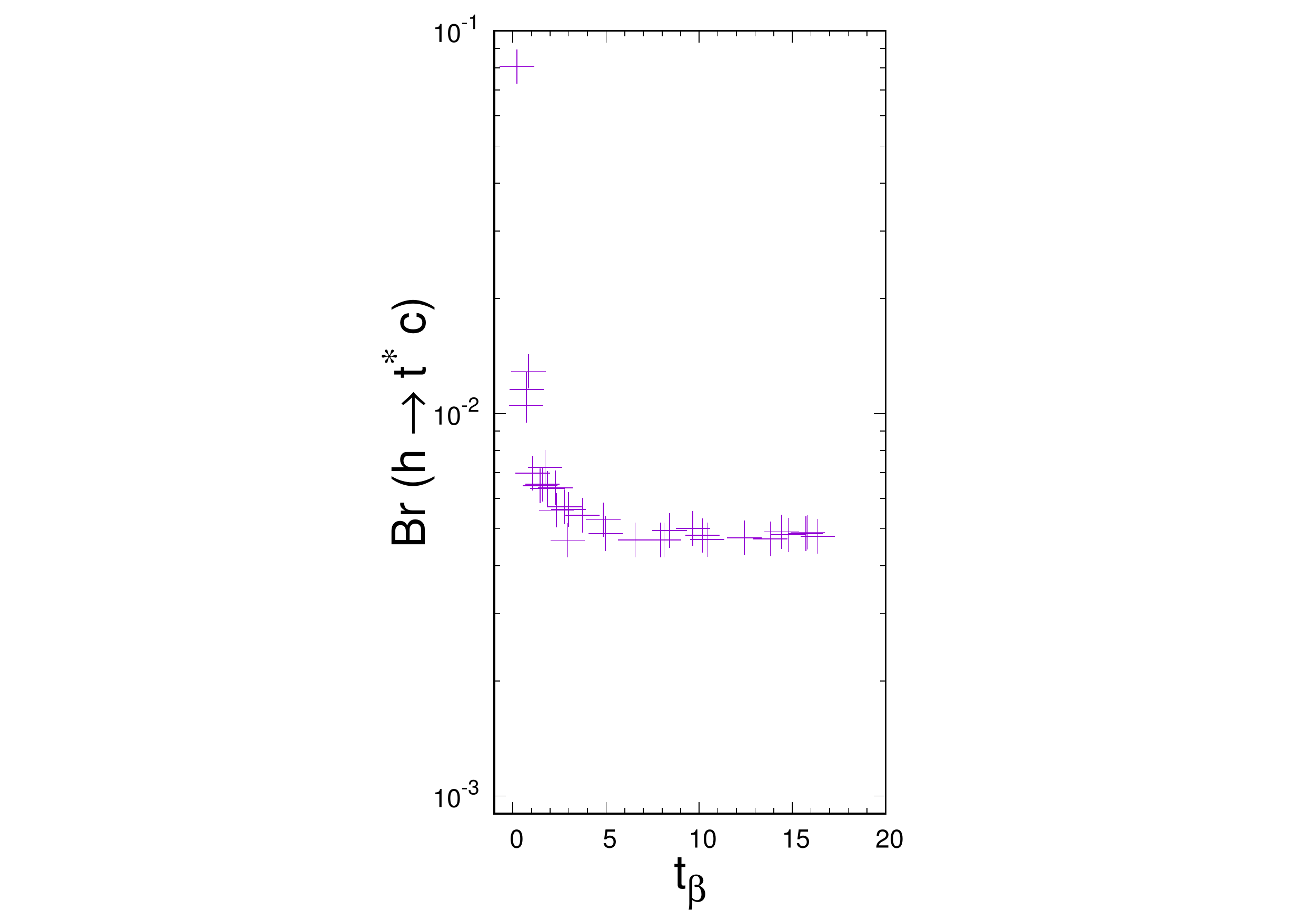}
\caption{\label{fig:Br_h-ts-c}
The ${\Br}$ for the process $h\to t^{*}_{} c$ that we 
have predicted.}
\end{figure}

If we consider the $W^{-}$ decays to 
$\nu_l^{} l^{}_{}$ then  
we obtain ${\Br}(h \to l^{} l^{} q \bar{q})$ of order $10^{-4},$ therefore
our results are interesting if we compare to the 
experimental results 
$\big({\Br}(h \to l l q \bar{q})\sim10^{-2}-10^{-3}\big)$ 
as is shown in ref. \cite{Dittmaier:2011ti} 
for $m_h = 125 \GeV.$
We found ${\Br}(h \to t^{*}_{}c)\sim 10^{-3}$ for the 
$1\lesssim t_\beta^{}\lesssim20$ (fig. \ref{fig:Br_h-ts-c}), 
which is a very interesting channel to explore 
in the LHC or the next generation of colliders.

Fig. \ref{fig:Br_h-ts-c} shows an interesting channel
($\Br(h\to t^* c)$ )
to probe new physics and we expect the next 
experimental results to test the THDM-III as the 
simplest SM extension.

\section{Conclusions
\label{sec:conclusions}}

We have explored the $h\to t^* c$ 
processes, and checked the high suppression 
considering FC  mediated by Higgs boson. 

Our exploration in the $\chi_{ij}^{u}-\chi_{ij}^{d}-$parameter space showed the allowed regions for different 
$t_\beta^{}$ values 
(see figs. \ref{label1}-\ref{label2}). 
Besides
we explore the different modes for 
Higgs decays to  (see figs. \ref{label3}-\ref{label4}).
Our results showed ${\Br}(h \to \mu \tau)\lesssim10^{-5}$ 
and  ${\Br}(h \to \gamma Z)\sim 10^{-6}$ as long 
as ${\Br}(b \to s \gamma) \lesssim10^{-4}$  as 
is shown in figs. \ref{label3}-\ref{label4}, those 
engaging values for exploring in LHC.
Besides we predicted ${\Br}(h \to t^*_{}c^{}_{})
\sim 10^{-3}$ for $1\lesssim t_\beta^{}\lesssim20$, 
this a feasible channel to explore 
our model in the LHC. 

{\ack 
I acknowledge support from 
CONACYT-SNI (Mexico).}

\section*{References}
\bibliography{mybilbio}

\providecommand{\newblock}{}
\begin{thebibliography}{10}
\expandafter\ifx\csname url\endcsname\relax
  \def\url#1{{\tt #1}}\fi
\expandafter\ifx\csname urlprefix\endcsname\relax\def\urlprefix{URL }\fi
\providecommand{\eprint}[2][]{\url{#2}}

\bibitem{Arroyo:2013tna}
Arroyo M, Diaz-Cruz J~L, Diaz E and Orduz-Ducuara J~A 2016 {\em Chin. Phys.\/}
  {\bf C40} 123103 (\textit{Preprint} \eprint{1306.2343})

\bibitem{DiazCruz:2012xc}
Diaz-Cruz L, Diaz-Furlong A, Gait\'an-Lozano R and Montes~de Oca J 2012
  (\textit{Preprint} \eprint{1203.6893})

\bibitem{TheATLAScollaboration:2013nia}
collaboration T~A (ATLAS) 2013  {ATLAS-CONF-2013-081}

\bibitem{Hahn:1998yk}
Hahn T and Perez-Victoria M 1999 {\em Comput. Phys. Commun.\/} {\bf 118}
  153--165 (\textit{Preprint} \eprint{hep-ph/9807565})

\bibitem{Hahn:2000kx}
Hahn T 2001 {\em Comput. Phys. Commun.\/} {\bf 140} 418--431 (\textit{Preprint}
  \eprint{hep-ph/0012260})

\bibitem{Staub:2015kfa}
Staub F 2015 {\em Adv. High Energy Phys.\/} {\bf 2015} 840780
  (\textit{Preprint} \eprint{1503.04200})

\bibitem{Gunion:2002zf}
Gunion J~F and Haber H~E 2003 {\em Phys. Rev.\/} {\bf D67} 075019
  (\textit{Preprint} \eprint{hep-ph/0207010})

\bibitem{DiazCruz:2004pj}
Diaz-Cruz J~L, Noriega-Papaqui R and Rosado A 2005 {\em Phys. Rev.\/} {\bf D71}
  015014 (\textit{Preprint} \eprint{hep-ph/0410391})

\bibitem{ElKhadra:2002wp}
El-Khadra A~X and Luke M 2002 {\em Ann. Rev. Nucl. Part. Sci.\/} {\bf 52}
  201--251 (\textit{Preprint} \eprint{hep-ph/0208114})

\bibitem{Gaitan:2017cfa}
Gait\'an R, Montes~de Oca J and Orduz-Ducuara J 2017 {\em Progress of
  Theoretical and Experimental Physics\/} {\bf 2017} 073B02

\bibitem{Dittmaier:2011ti}
Dittmaier S {\em et~al.\/} (LHC Higgs Cross Section Working Group) 2011
  (\textit{Preprint} \eprint{1101.0593})

\end{thebibliography}

\end{document}